# Giant room-temperature barocaloric effects in PDMS rubber at low pressures

A. M. G. Carvalho[1,*], W. Imamura[1,2], E. O. Usuda[1], N. M. Bom[1]

[1]Laboratório Nacional de Luz Síncrotron (LNLS), Centro Nacional de Pesquisa em Energia e Materiais (CNPEM), CEP 13083-970, Campinas, São Paulo, Brazil.

[2]Faculdade de Engenharia Mecânica, UNICAMP, CEP 13083-860, Campinas, SP, Brazil.

## ABSTRACT

The barocaloric effect is still an incipient scientific topic, but it has been attracting an increasing attention in the last years due to the promising perspectives for its application in alternative cooling devices. Here, we present giant values of barocaloric entropy change and temperature change induced by low pressures in PDMS elastomer around room temperature. Adiabatic temperature changes of 12.0 K and 28.5 K were directly measured for pressure changes of 173 MPa and 390 MPa, respectively, associated with large normalized temperature changes (~70 K GPa$^{-1}$). From adiabatic temperature change data, we obtained entropy change values larger than 140 J kg$^{-1}$ K$^{-1}$. We found barocaloric effect values that exceed those previously reported for any promising barocaloric materials from direct measurements of temperature change around room temperature. Our results stimulate the study of the barocaloric effect in elastomeric polymers and broaden the pathway to use this effect in solid-state cooling technologies.

**Introduction**

In the beginning of the nineteenth century, John Gough detected and described the heating of natural rubber when rapidly stretched [1]. This effect was further studied by Joule, reporting temperature changes induced by uniaxial stress in different materials, including rubbers, metals, and woods [2]. In fact, both scientists described what we designate elastocaloric effect ($\sigma_e$-CE), the first *i*-caloric effect ever reported. *i*-caloric effects ("*i*" stands for intensive thermodynamics variables) can be characterized by the adiabatic temperature change ($\Delta T_S$) and the isothermal entropy change ($\Delta S_T$) induced by an external field applied on a material.

Besides the magnetocaloric effect (*h*-CE) and the electrocaloric effect (*e*-CE), Lord Kelvin [3] also predicted the barocaloric effect ($\sigma_b$-CE), which should be driven by isotropic stress variations. As well as *h*-CE and *e*-CE, the $\sigma_b$-CE may be used in solid-state cooling devices through a cooling cycle, as illustrated in Fig. 1. Both $\sigma_b$-CE and $\sigma_e$-CE are facets of the mechanocaloric effect ($\sigma$-CE). The research into *i*-caloric effects has blossomed in the last decades, as a consequence of the experimental demonstration of the giant *h*-CE [4] and the giant *e*-CE [5], leading to significant advances in materials and prototypes. Despite the interesting results obtained by stretching natural rubber (NR) and other polymers in the 1940 decade [6,7] or pressing glassy polymers, such as poly (methyl methacrylate) [8], the $\sigma$-CE is the least studied *i*-caloric effect, contrasting the large number of papers on *h*-CE and *e*-CE. Nevertheless, recently, there was a resurgence of $\sigma$-CE due the practicality of applying mechanical stress in comparison to magnetic or electrical fields.

Giant values of $\sigma_e$-CE and $\sigma_b$-CE around room temperature were reported in shape-memory alloys (SMAs) based on Cu-Zn-Al ($\Delta S_T$ = -22 J kg$^{-1}$ K$^{-1}$ at 295 K, for $\Delta\sigma$ = 143 MPa) [9] and Ni-Mn-In ($\Delta S_T$ = -24.4 J kg$^{-1}$ K$^{-1}$ and $\Delta T_S$ = 4.5 K at 293 K, for $\Delta\sigma$ = 260 MPa) [10], respectively. Other examples of giant $\sigma$-CE around room temperature are observed in Ni$_{46}$Mn$_{38}$Sb$_{12}$Co$_4$, Fe$_{49}$Rh$_{51}$ and BaTiO$_3$ [11–13]. Below room temperature, (NH$_4$)$_2$SO$_4$ presents promising barocaloric properties ($\Delta S_T$ = -60 J kg$^{-1}$ K$^{-1}$ for $\Delta\sigma$ = 100 MPa) [14]. Regarding organic-inorganic materials, large $\sigma_b$-CE values were obtained for several families of hybrid perovskites [15]. The promising mechanocaloric potential of ferroics stimulated the recent development of a regenerative elastocaloric heat pump made of a Ni-Ti alloy [16].

Alternatively to ferroics, elastomeric polymers have also attracted some attention regarding the σ-CE [17–23]. Elastomers have shown to be particularly suitable for mechanocaloric applications, since they present good fatigue properties combined with high caloric potential [24]. Moreover, all elastomers can act as mechanocaloric materials due to the contribution of polymer chains rearrangement to the σ-CE, not depending on phase transitions (differently from ferroics).

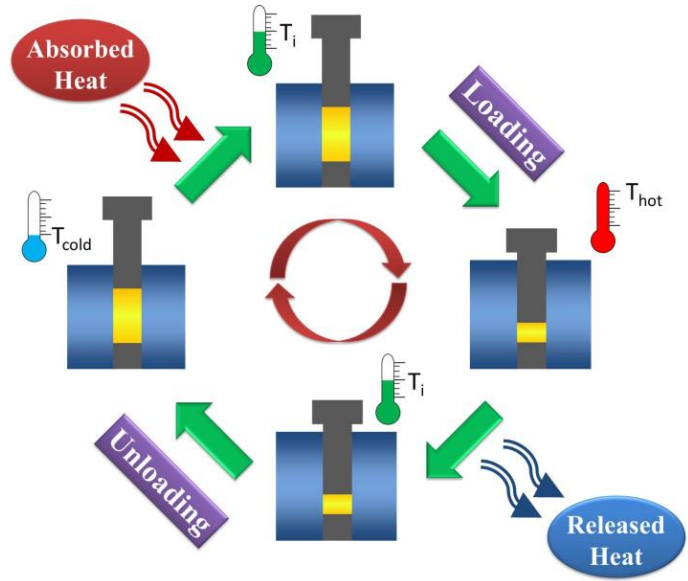

**Figure 1.** Scheme of a four-stage barocaloric cycle based on loading/unloading process. The yellow rectangles represent the solid refrigerant during the cycles. As a first step, the material is adiabatically compressed, increasing the sample temperature ($T_{hot}$). Then, while the applied pressure is kept constant, the heat gradually flows out from the sample to the heat sink and the temperature decreases down to initial temperature ($T_i$). In the third step, the load is adiabatically released and the sample is cooled down ($T_{cold}$). Finally, the sample absorbs heat from the surroundings and returns to $T_i$.

In this context, the polydimethylsiloxane rubber (PDMS) appears as a potential mechanocaloric material. PDMS is a well-studied polymer for several applications, such as medicine, food industry, toxicity tests, microfabrication [25]. From a physico-chemical point of view, PDMS is an elastomer, i.e., composed of long-chain molecules with freely rotating links, weak secondary forces between the molecules, and cross-links able to form a three-dimensional network [26]. Moreover, PDMS is also amorphous around room temperature. Combining these particularities with confined compression process, we can hypothesize that the entropy or the temperature can be changed more easily in PDMS than in other non-rubber-like solids (e.g., SMAs). Therefore, we have systematically investigated the $\sigma_b$-CE in PDMS around room temperature at relatively low pressures. Our experiments reveal that PDMS exhibits giant temperature and entropy variations, presenting a great potential as a refrigerant in barocaloric solid-state cooling devices.

**Experimental**

The PDMS samples were prepared from the commercial components supplied by Dow Corning®. A pre-polymer base and a curing agent (Sylgard 184) were mixed together at the recommended mass proportion of 10:1. To avoid air bubbles, the mixture was put in low vacuum for about 45 minutes. We used two metallic cylinders with two diameters (8 and 12 mm) as a mold for the samples. The mixture was poured into the mold over a glass plate until it was completely filled and then placed on a hot plate (368 K) for 50 minutes. We made two samples with the following dimensions: 12 mm (diameter) and 17 mm (length); 8 mm (diameter) and 20 mm (length). The densities of the samples are 1026(3) and 1030(7) kg m$^{-3}$, respectively. In the experiments performed in pressures up to 173 MPa, we used the same PDMS sample with 12 mm of diameter. For experiments above 173 MPa, the same 8-mm-diameter sample was used. We characterized the 12-mm-diameter sample via Fourier transform infrared spectroscopy (FTIR) from 450 to 4000 cm$^{-1}$, with a fixed step of 2 cm$^{-1}$ (Fig. 2), using a FTIR spectrometer from PerkinElmer® (model Spectrum Two).

The experimental setup consists of a cylinder carbon-steel chamber enveloped by a copper coil, which water or liquid nitrogen flows inside for cooling/heating. For temperatures above 280 K, a thermostatic bath (TE 184, Tecnal) was used to pump water in the copper coil. For temperatures below 280 K, we used liquid nitrogen to refrigerate the sample. A set of two resistors (NP 38899, HG Resistências) placed in the proper holes in the chamber helps control precisely the temperature when used together with the liquid nitrogen. Two variations of the chamber were used: one with a 12-mm-diameter cylindrical hole and another one with an 8-mm-diameter cylindrical hole. Pistons with the respective diameters move through the cylindrical holes in the center of the chamber, where the sample is placed in. Below the sample, there is a bottom closure holding the sample inside the chamber and guiding a thermocouple, whose tip is placed inside the sample to measure its temperature in real time. Uniaxial load at the piston is applied by a manual hydraulic press (P15500, Bovenau); the applied tensions are isostatic for confined elastomers (see discussion in Appendix). Underneath the system, a load cell (3101C, ALFA Instrumentos) measures the contact force. Sample displacement is measured by a precise linear length gauge (METRO 2500, Heidenhain Co), with the help of a rod attached to the piston. Temperature data are collected and controlled (when using the resistors) by Model 335, Lake Shore Cryotronics. A schematic view of the system is reported by Bom et al. [29].

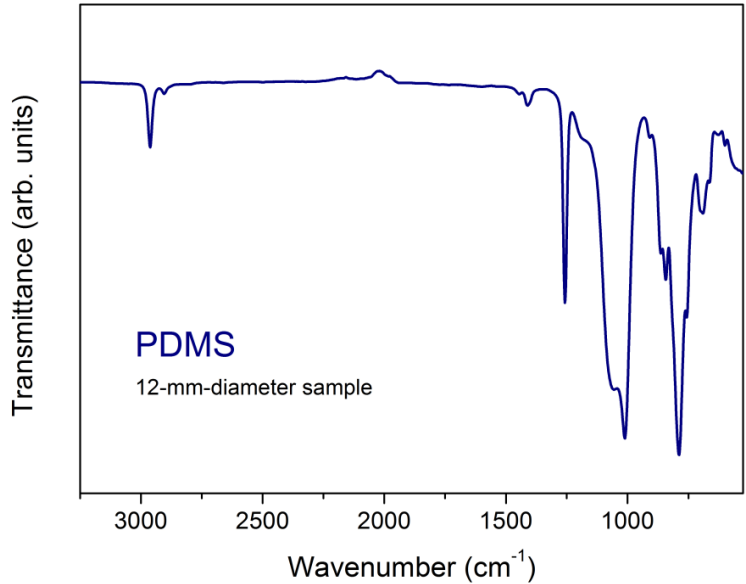

**Figure 2.** FTIR spectrum of the 12-mm-diameter sample of PDMS at room temperature. The absorption bands are in agreement to those previously reported [27,28].

Direct $\Delta T_S$ were obtained by applying pressure very close to the adiabatic condition. When the temperature in the sample is stable, a compressive stress (maximum values in the range of 26.0 – 390 MPa) is applied rapidly, resulting in an immediate increase in temperature. The load is kept constant until the temperature decreases down to the initial temperature. Finally, the stress is released adiabatically, causing an abrupt decrease in the sample temperature.

Strain *vs.* temperature experiments were performed at isobaric processes, i.e., the temperature was reduced under a constant pressure at the 0.7 – 130 MPa range. Firstly, we set a temperature of 328 K in the sample. Then, a compressive stress is applied and kept constant. Temperature is varied continuously between ~328 – 283 K, performing the following cycle: 328 K → 283 K → 328 K → 283 K.

**Results and discussion**

Adiabatic temperature change and isothermal entropy change

Firstly, we measured $\Delta T_S$ for different pressure variations (Fig. 3a). We observe a thermal giant $\sigma_b$-CE of 27.7 K at ~283 K, for $\Delta\sigma$ = 390 MPa (compression), and $\Delta T_S$ = -28.5 K, at the same initial temperature, for $\Delta\sigma$ = -390 MPa (decompression). The difference between $\Delta T_S$ values in compression and decompression processes, for the highest pressure variations, is probably due to the fact that the decompression process is closer to the adiabatic condition than the compression process. Taking the process 0 MPa → 390 MPa → 0 MPa as an example, the temperature relaxation during compression has a time constant of $\tau_{comp}$ = 19.1 s, while the time constant is $\tau_{decomp}$ = 24.6 s for decompression. Then, if $\tau_{comp} < \tau_{decomp}$ and the sample specific heat does not change significantly during the process, the heat transfer rate for compression is larger than decompression.

Fig. 3b displays the $\Delta T_S$ as function of temperature in decompression process within the 26.0(5) – 390(12) MPa pressure range. At 303 K, for example, it was obtained a $\Delta T_S$ of -12.1 K for just 173(3) MPa, higher than observed for vulcanized natural rubber (V-NR) under the same conditions ($\Delta T_S$ = -10.5 K at 303 K, for $\Delta\sigma$ = -173(3) MPa) [23]. Furthermore, PDMS at low temperatures (223 and 243 K) was also analyzed for $\Delta\sigma$ of 43.4(9) and 173(3) MPa, showing high $\Delta T_S$ even below room temperature. This behavior is interesting since $\Delta T_S$ does not change significantly in a wide temperature range (i.e., $\Delta T_S$ does not decrease abruptly as in magnetocaloric compounds [30] or SMAs [31–33]). It is worth noticing that $\Delta T_S$ curves present some temperature dependence only at the highest pressures (273 and 390 MPa). This could be explained by the fact that an 8-mm-diameter sample was used at this pressure range, different from that measured at lower pressures (12 mm of diameter). Although the same synthesis process was employed for both samples, slight variations in the final product may occur, which can exert some influence on caloric parameters.

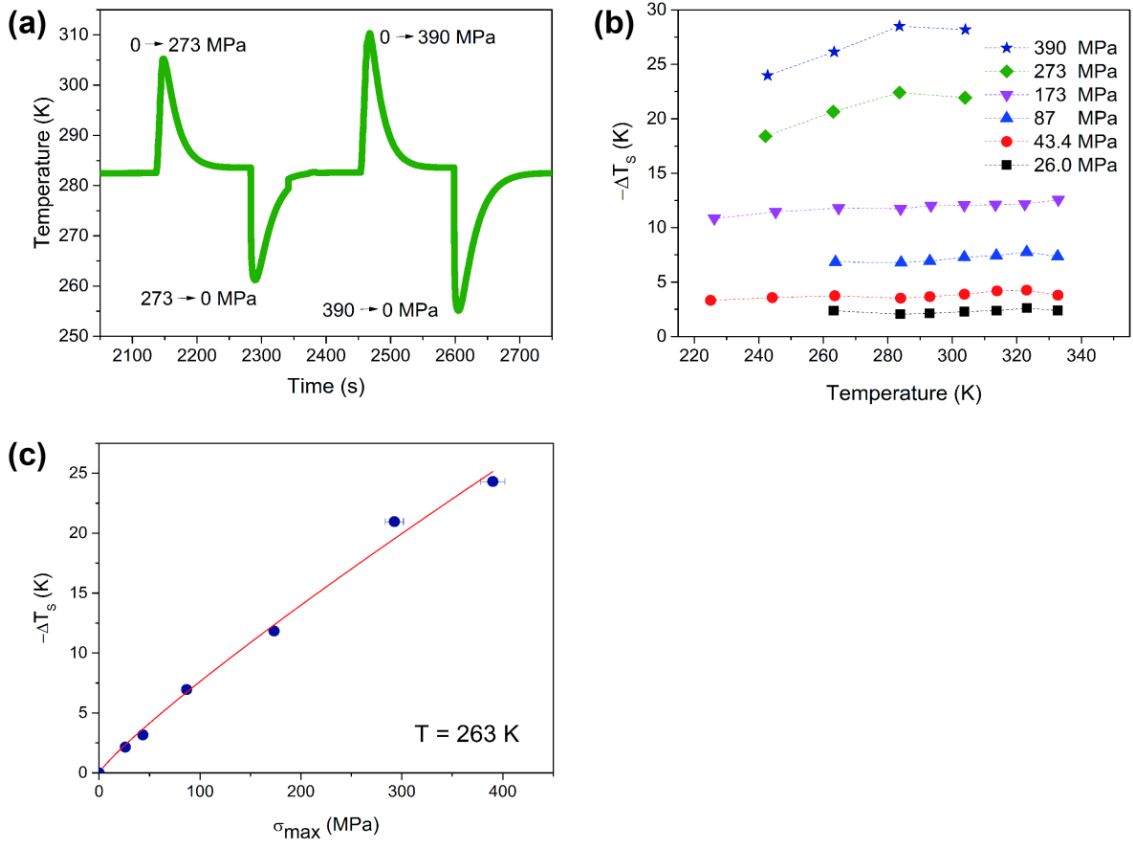

**Figure 3.** (a) Temperature *vs.* time for PDMS rubber at initial temperature of ~283 K; the peaks/valleys are related to the adiabatic temperature change ($\Delta T_S$) when the pressures of 273(8) and 390(12) MPa are applied/released. (b) $\Delta T_S$ *vs.* initial temperature for decompression process at different pressures (26.0(5), 43.4(9), 87(2), 173(3), 273(8) and 390(12) MPa); the dotted lines connecting the symbols are guides for the eyes. (c) $\Delta T_S$ *vs.* released pressure for initial temperature of 263 K; the circles are experimental data, and the line is the fit using Eq. 1; we estimate an error of 2% for pressures up to 173 MPa and 3% above this pressure.

For several materials presenting *i*-caloric effects, the maximum magnitude of the $\Delta T_S$ increases as a function of the maximum applied field, following a power law of the type $\Delta T_S(T,i) = ai^n$. For example, the external field in the *h*-CE is the magnetic field (H), and $n = 2/3$ around the magnetic transition. In that case, $\Delta T_S$ and $\Delta S_T$ are proportional to $H_{max}^{2/3}$, which is predicted from the Mean Field Theory [34]. For *e*-CE, a power law was reported for a ferroelectric polymer, for which $n = 2$ (i.e., $\Delta T_S \propto E_{max}^2$) at low electric fields (E), and $n = 2/3$ at higher E (i.e., $\Delta T_S \propto E_{max}^{2/3}$) [35]. Following this concept, Usuda et al. [23] also proposed a power law for $\sigma_b$-CE:

$$|\Delta T_S(T, \sigma_{max})| = a\sigma_{max}^n, \quad (1)$$

where $a$ is the constant of proportionality, and $\sigma_{max}$ is the maximum value of the released pressure. The fitted model for PDMS at 263 K is shown in Fig. 3c. In this case, $a = 62(5)$ K GPa$^{-n}$, and $n = 0.90(6)$. For other temperatures, $a$ and $n$ parameters are shown in Table A1, in appendix. As an informative comparison, the same parameters for V-NR [23] at 283 K are $a = 51(3)$ K GPa$^{-n}$ and $n = 0.94(3)$. Either for PDMS or V-NR, $\Delta T_S$ *vs.* $\sigma_{max}$ could be fitted by a power law. For PDMS, the parameters $a$ and $n$ slightly increase at higher temperatures.

We have indirectly determined the $\Delta S_T$ from strain as a function of temperature experimental curves (Figs A1a-b, Appendix), through the following Maxwell's relation [23,34,35]:

$$\Delta S_T(T, \Delta\sigma) = -\frac{1}{\rho_0}\int_{\sigma_1}^{\sigma_2}\left(\frac{\partial \varepsilon}{\partial T}\right)_\sigma d\sigma, \quad (2)$$

where $\rho_0$ is the density of the sample at atmospheric pressure ($\sigma_0$) and ambient temperature (T$_0$ = 293 K); $\sigma_1$ is the initial pressure ($\sigma_1 \approx \sigma_0$); $\sigma_2$ is the final pressure; $\varepsilon$ is the strain, defined as $\varepsilon(\sigma, T) \equiv (l_{\sigma,T} - l_0)/l_0$, where $l_{\sigma,T}$ is the final length of the sample at $\sigma$, for each temperature T, and $l_0$ is its initial length at $\sigma_0$ and T$_0$. The curves for $\Delta S_T$ as function of temperature, calculated by Eq. 2, are displayed in Fig. 4a. At 290 K, we obtained a giant $\sigma_b$-CE of -53(5) J kg$^{-1}$ K$^{-1}$ for merely $\Delta\sigma = 130(3)$ MPa. This leads to a normalized entropy change ($\left|\Delta S_T/\Delta\sigma\right|$) of 0.41(5) kJ kg$^{-1}$ K$^{-1}$ GPa$^{-1}$. For $\Delta\sigma = 87(2)$ MPa, the entropy change is lower, -42(4) J kg$^{-1}$ K$^{-1}$, but the normalized entropy change keeps its high value: 0.48(6) kJ kg$^{-1}$ K$^{-1}$ GPa$^{-1}$. For the highest pressure changes, $\Delta S_T$ shows a tendency of increasing at temperatures below 290 K.

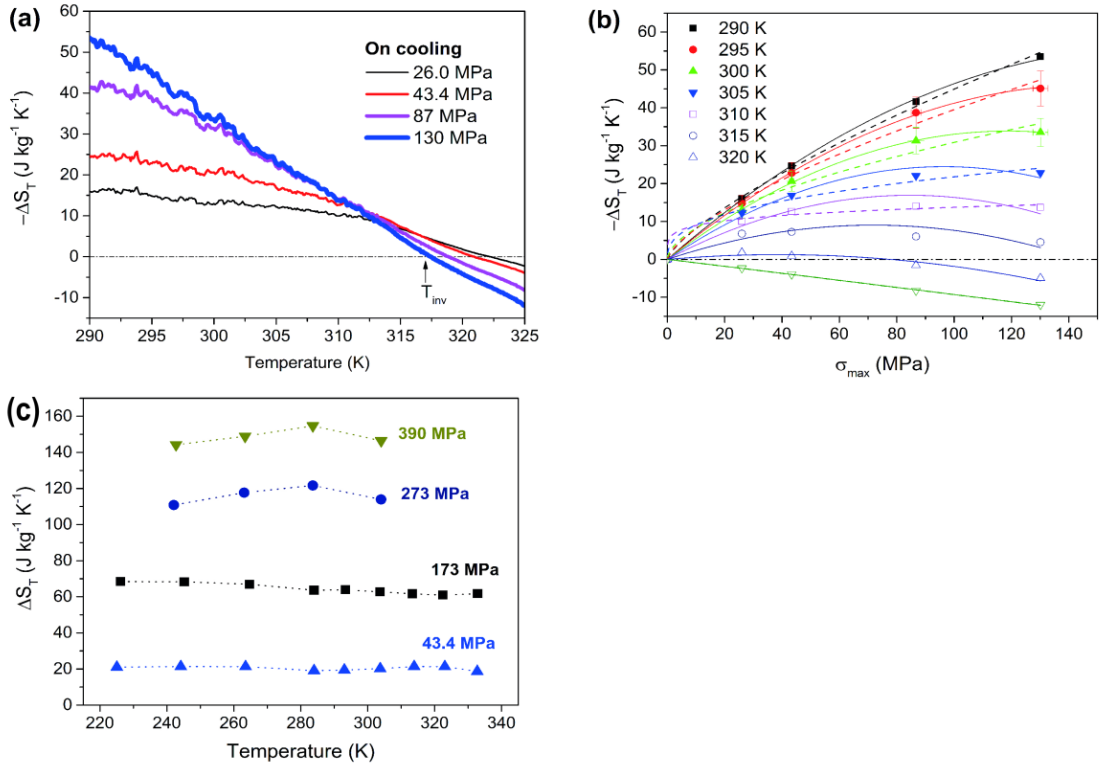

**Figure 4.** (a) Isothermal entropy change ($\Delta S_T$) as a function of temperature for $\Delta\sigma$ = 26.0(5), 43.4(9), 87(2) and 130(3) MPa, obtained from $\varepsilon$ *vs*. *T* on cooling process by Eq. 2; the horizontal dash-dotted line at zero is a guide for the eyes; $T_{inv}$ is the temperature that marks the inversion of the $\sigma_b$-CE (for 130(3) MPa as an example). (b) $\Delta S_T$ on cooling process *vs*. maximum applied pressure for different temperatures; the dotted lines are the fits using Eq. 3; the solid lines are the fits using Eq. 4; the horizontal dash-dot line at zero is a guide for the eyes to separate the ordinary $\sigma_b$-CE from the inverse $\sigma_b$-CE; the error bars (2% for $\sigma_{max}$ and 1 J kg$^{-1}$ K$^{-1}$ + 8% for $\Delta S_T$) were only shown for 295 and 300 K data for the sake of clarity; (c) $\Delta S_T$ as a function of temperature for $\Delta\sigma$ = 43.4(9), 173(3), 273(8) and 390(12) MPa, obtained from $\Delta T_S$ data in Fig. 3b by Eq. 5; the dotted lines connecting the symbols are guides for the eyes.

Here, we figure that an entropy state of an amorphous elastomer can be directly associated with molecular motion related to vibrations, rotations and small relative displacements among the polymer chains. Under pressure (and without transitions), the chains are compressed and the molecular motion is reduced, following a decrease in free volume (the total volume minus the volume occupied by chains), resulting in a lower entropy state. Within certain temperature ranges, the reductions in free volume and molecular motion can be very pronounced, resulting in large entropy changes. Moreover, PDMS also exhibits inverse $\sigma_b$-CE (inverse $\Delta S_T$) at higher temperatures (e.g., for $\Delta\sigma$ = 130(3) MPa, above $T_{inv} \approx$ 317 K). The temperature $T_{inv}$, which marks the inversion of the $\sigma_b$-CE in PDMS, shifts to higher values as $\Delta\sigma$ decreases. The inverse $\Delta S_T$ is due to the negative derivative observed in the higher temperature region of ε vs. temperature curves (Fig. A1c, Appendix). The observation of an inverse $\sigma_b$-CE in -$\Delta S_T$ vs. T curves (Fig. 4a) may seem inconsistent with the -$\Delta T_S$ vs. T curves (Fig. 3b), where no anomalous behaviors are verified. However, $\Delta T_S$ and $\Delta S_T$ of PDMS were obtained from two different thermodynamic processes (see Experimental section). There is some irreversibility in ε measurements, since we observe different paths on cooling and on heating measurements (Fig. A2, Appendix). Besides, even in a cycle (ε vs. temperature), the material does not return to its initial state. Thus, $\Delta S_T$ is affected by this irreversibility, analogously to what is observed in magnetocaloric materials, which can manifest ordinary or inverse $\Delta S_T$ depending on the measurement protocols [36]. Therefore, the behavior of the directly obtained $\Delta T_S$ curves and the $\Delta S_T$ curves from ε vs. T data may be very different.

In Fig. 4b, we show $\Delta S_T$ as a function of the maximum applied pressure ($\sigma_{max}$) for different temperatures. We fitted the ordinary $\sigma_b$-CE for different temperatures also using a power law as following:

$$-\Delta S_T(T, \sigma_{max}) = b\sigma_{max}^m, \qquad (3)$$

where *b* is the constant of proportionality, and *m* is the power-law exponent. As examples: for 290 K, $b$ = 0.25(2) kJ kg$^{-1}$ K$^{-1}$ GPa$^{-m}$ and $m$ = 0.74(3); and, for 295 K, $b$ = 0.19(3) kJ kg$^{-1}$ K$^{-1}$ GPa$^{-m}$ and $m$ = 0.69(7) (all fitting parameters are shown in Table A2, Appendix). It is easy to see that *b* and *m* decrease when the temperature increases for ordinary $\sigma_b$-CE. Furthermore, the power law works while $\left(\frac{d\Delta S_T}{d\sigma_{max}}\right)_T$ does not change its sign for the entire range of pressure (i.e., the power law fails for 315 and 320 K). The behavior of the power-law parameters for $\Delta S_T$ (*b* and *m*) as well as the parameters for $\Delta T_S$ (*a* and *n*) need further investigation to be better understood.

We also fitted -$\Delta S_T$ vs. $\sigma_{max}$ for different temperatures using the following equation:

$$-\Delta S_T(T, \sigma_{max}) = a_1(T)\sigma_{max} + a_2(T)\sigma_{max}^2, \qquad (4)$$

derived from a modified Landau's theory of elasticity (see Appendix), where $a_1$ and $a_2$ are parameters to fit (Table A2, Appendix). This quadratic model fits better the experimental data than the power-law model, and it is able to fit properly the isotherms of 315 and 320 K as well.

Another approach to calculate $\Delta S_T$ is through the equation below [37]:

$$\Delta S_T(T, \Delta\sigma) = -\frac{c_p(T)}{T}\Delta T_S(T, \Delta\sigma), \qquad (5)$$

where $c_p (T)$ is the specific heat as a function of temperature (for PDMS, see Ref. [38]). Since Eq. 5 is valid only far from transitions [37], and our PDMS does not present transitions within the measured temperature range, the calculation of $\Delta S_T$ vs. T curves from direct $\Delta T_S$ data is a good approximation. In this case, it is possible to evaluate the $\Delta S_T$ for pressures up to $\Delta\sigma = 390$ MPa, according to the experimental conditions of $\Delta T_S$ collection (Fig. 3b). $\Delta S_T$ values obtained from this method are displayed in Fig. 4c. One can observe that the qualitative behavior of these curves significantly differs from those in Fig. 4a (from ε vs. T data), which exhibit a strong dependence on temperature; these two protocols for $\Delta S_T$ show a similar $\Delta S_T \approx 20$ J kg$^{-1}$ K$^{-1}$ in the temperature range of 290 – 300 K, for $\Delta\sigma = 43.4$ MPa. $\Delta S_T$ curves displayed in Figure 3c follow the trend verified in $\Delta T_S$ dataset (Fig. 3b). Additionally, we used Eq. 5 to estimate indirect $\Delta T_S$ vs. T curves (Fig. A3, Appendix) from $\Delta S_T$ (Fig. 4a) obtained from ε vs. T data; in this case, indirect and direct $\Delta T_S$ are similar between 290 and 300 K, for $\Delta\sigma = 26.0$, 43.4 and 87 MPa, reaching $\Delta T_S \approx 2$, 4 and 8 K, respectively.

Performance parameters

Our obtained values of $\Delta T_S$ and $\Delta S_T$ for PDMS are striking not only due to their magnitude, but also because they were observed at relatively low applied pressures and strains. In tractive $\sigma_e$-CE of elastomers at low stresses, $\Delta T_S$ values greater than 10 K require large strain amplitudes of 400 – 700% [6,17,39]. On the other hand, giant thermal $\sigma_b$-CE values (> 20 K) in ferroics are exhibited in small strain amplitudes, but often require high pressures (several hundreds of MPa) [32]. Thus, tractive $\sigma_e$-CE in polymers and $\sigma_b$-CE in ferroics present significant drawbacks concerning applications in cooling devices. Now, considering $\sigma_b$-CE in elastomers, such as PDMS (giant $\Delta T_S$, giant $\Delta S_T$, $\sigma < 300$ MPa and $|\varepsilon| < 20\%$), this scenario seems to change.

In order to compare the barocaloric properties of PDMS around room temperature with different relevant barocaloric materials from the literature [12,14,20,40–43], we present the *normalized temperature change* ($|\Delta T_S/\Delta\sigma|$) as a function of temperature (Fig. 5a) and as a function of $\Delta T_S$ (Fig. 5b). A remarkable $|\Delta T_S/\Delta\sigma|$ of ~70 K GPa$^{-1}$ was obtained for PDMS, which remains practically constant within a large temperature range. Despite the fact that a few materials exceed this value under particular conditions, PDMS presents the highest $|\Delta T_S/\Delta\sigma|$ for $\Delta T_S > 10$ K. Moreover, it is relevant to stress that our $\Delta T_S$ values of PDMS were measured directly, on the contrary of most barocaloric materials reported so far.

We also calculated the *normalized refrigerant capacity* (NRC) as a function of the temperature difference between hot reservoir and cold reservoir ($\Delta T_{h-c} \equiv T_{hot} - T_{cold}$). We define the normalized refrigerant capacity for mechanocaloric effect as:

$$NRC(\Delta T_{h-c}, \Delta\sigma) = \left|\frac{1}{\Delta\sigma}\int_{T_{cold}}^{T_{hot}} \Delta S_T(T,\Delta\sigma)dT\right|, \qquad (6)$$

For PDMS (Fig. 5c), we fixed the hot reservoir at 315 K. It is noteworthy that NRC sharply increases as function of $\Delta T_{h-c}$, surpassing 5 and 9 kJ kg$^{-1}$ GPa$^{-1}$ for $\Delta S_T$ obtained from $\varepsilon$ *vs.* T data and from $\Delta T_S$ *vs.* T, respectively. Finally, the absolute energy efficiency of a caloric material can be evaluated by calculating the *coefficient of performance* (COP). This parameter can be defined as $COP(T,\Delta\sigma) = |Q|/W$, where $Q = T\,\Delta S_T$ and $W = \rho_0^{-1}\int_{\varepsilon_1}^{\varepsilon_2}\sigma\,d\varepsilon$. The calculated COP values for PDMS at 290 K, for $\Delta\sigma$ = 130 and 87 MPa, are 4.7(7) and 8(1), respectively (using $\Delta S_T$ values from Fig. 4a). Considering the maximum theoretical efficiency (i.e., Carnot efficiency) operating at $T_{hot}$ = 315 K and $T_{cold}$ = 290 K, the value of COP$_{Carnot}$ = $T_{cold}/(T_{hot} - T_{cold})$ is 11.6. So, the relative COP ($\eta$ = COP$_{PDMS}$/COP$_{Carnot}$) is 41(6)% and 66(8)% for $\Delta\sigma$ = 130 and 87 MPa, respectively.

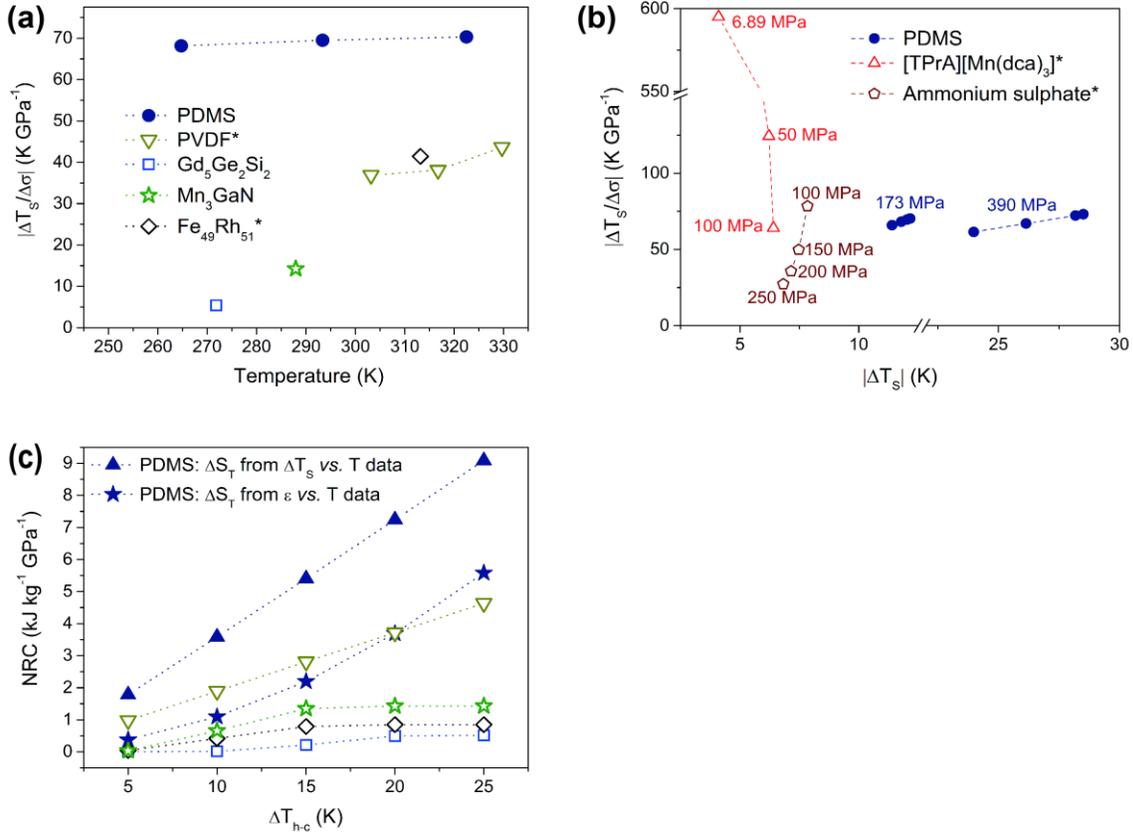

**Figure 5.** Barocaloric properties for materials with large or giant $\sigma_b$-CE around room temperature (250 - 330 K). (a) Normalized temperature change ($|\Delta T_S/\Delta\sigma|$): PDMS ($|\Delta\sigma|$ = 173 MPa); PVDF ($|\Delta\sigma|$ = 150 MPa) [20]; $Gd_5Ge_2Si_2$ (maximum reported for $|\Delta\sigma|$ = 200 MPa) [40]; $Mn_3GaN$ (maximum reported for $|\Delta\sigma|$ = 93 MPa) [36]; $Fe_{49}Rh_{51}$ (maximum reported for $|\Delta\sigma|$ = 250 MPa) [42]. (b) $|\Delta T_S/\Delta\sigma|$ as a function of absolute temperature change ($|\Delta T_S|$) for PDMS, [TPrA][Mn(dca)$_3$] [43], and $(NH_4)_2SO_4$ [14], at different $|\Delta\sigma|$ (values obtained from maximum $|\Delta S_T|$ for [TPrA][Mn(dca)$_3$] and $(NH_4)_2SO_4$). (c) Normalized refrigerant capacity (NRC) as a function of $\Delta T_{h-c} \equiv T_{hot} - T_{cold}$ (temperature difference between hot reservoir and cold reservoir): PDMS ($T_{hot}$ = 315 K, closed triangles: $\Delta S_T$ from $\Delta T_S$ *vs.* T data for $|\Delta\sigma|$ = 173 MPa, closed stars: $\Delta S_T$ from ε *vs.* T data for $|\Delta\sigma|$ = 130 MPa); PVDF ($T_{hot}$ = 330 K, $|\Delta\sigma|$ = 150 MPa) [20]; $Gd_5Ge_2Si_2$ ($T_{hot}$ = 275 K, $|\Delta\sigma|$ = 150 MPa) [40]; $Mn_3GaN$ ($T_{hot}$ = 295 K, $|\Delta\sigma|$ = 139 MPa) [41]; $Fe_{49}Rh_{51}$ ($T_{hot}$ = 325 K, $|\Delta\sigma|$ = 160 MPa) [12]. *Indirect determination.

**Conclusions**

In summary, we presented outstanding results concerning the barocaloric properties of PDMS rubber. The pressure-induced temperature changes are huge (e.g., at ~283 K, $\Delta T_S$ = -28.5 K, for $\Delta\sigma$ = -390 MPa; or $\Delta T_S$ = -22.4 K, for $\Delta\sigma$ = -273 MPa), as well as the isothermal entropy changes ($\Delta S_T$ > 140 J kg$^{-1}$ K$^{-1}$, for 290 K and $\Delta\sigma$ = 390 MPa). Regarding temperature changes, the barocaloric effect values presented here surpass those for any other barocaloric material in literature obtained from direct measurements around room temperature. Considering normalized parameters, PDMS exhibits the normalized temperature change $|\Delta T_S/\Delta\sigma| \approx$ 70 K GPa$^{-1}$ within the temperature range of 265 – 322 K, and the normalized refrigerant capacity NRC are higher than 9 kJ kg$^{-1}$ GPa$^{-1}$ ($\Delta T_{h-c}$ = 25 K). Therefore, the striking barocaloric effects of PDMS observed at low pressures and low strains open a promising road towards solid-state cooling based on elastomeric polymers submitted to confined compression.

**Conflicts of interest**

The authors declare that there are no conflicts of interest.

**Acknowledgements**

The authors acknowledge financial support from FAPESP (project number 2012/03480-0), CNPq, CAPES, LNLS and CNPEM. The authors also thank Francesco G. Carotti and Maria Helena O. Piazzetta for the support in the preparation of PDMS samples and Rafael O. Martins for technical support.

**Appendix**

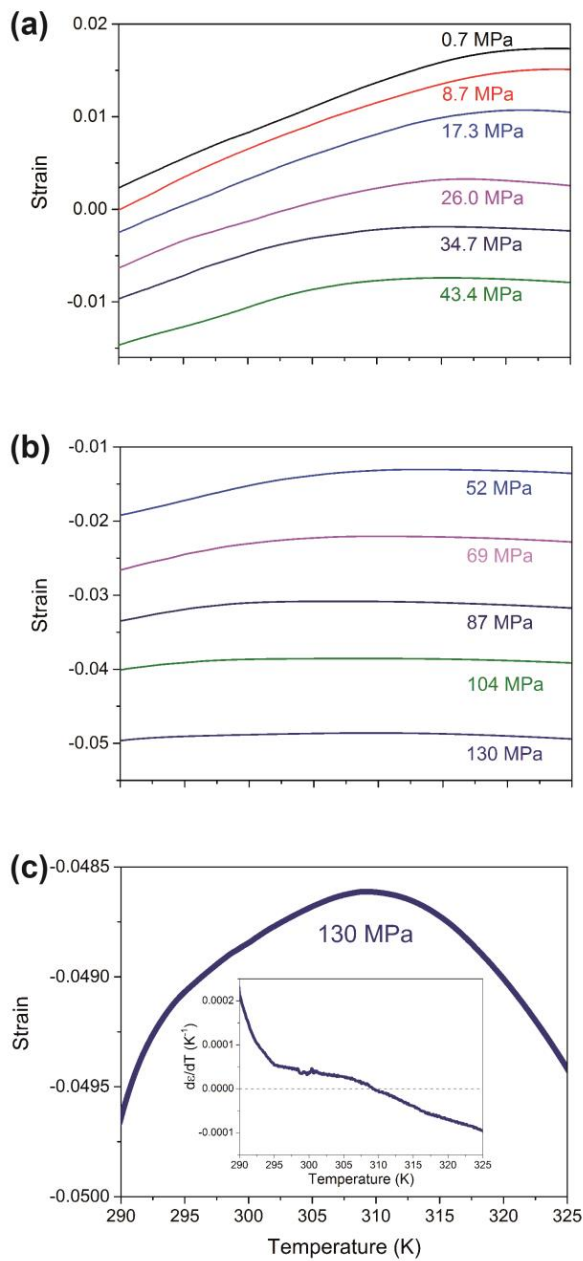

**Figure A1.** (a) Strain *vs.* temperature curves for PDMS at constant pressures of 0.7(1), 8.7(2), 17.3(3), 26.0(5), 34.7(7), 43.4(9) MPa measured on cooling. (b) Strain *vs.* temperature curves for PDMS at constant pressures of 52(1), 69(1), 87(2), 104(2) and 130(3) MPa measured on cooling; together with the curves of Fig A1a, they were used to calculate the isothermal entropy change displayed in Fig. 4a (main text). (c) Strain *vs.* temperature curve at 130 MPa; the inset shows the sign change of the derivative $\left(\frac{\partial \varepsilon}{\partial T}\right)_\sigma$.

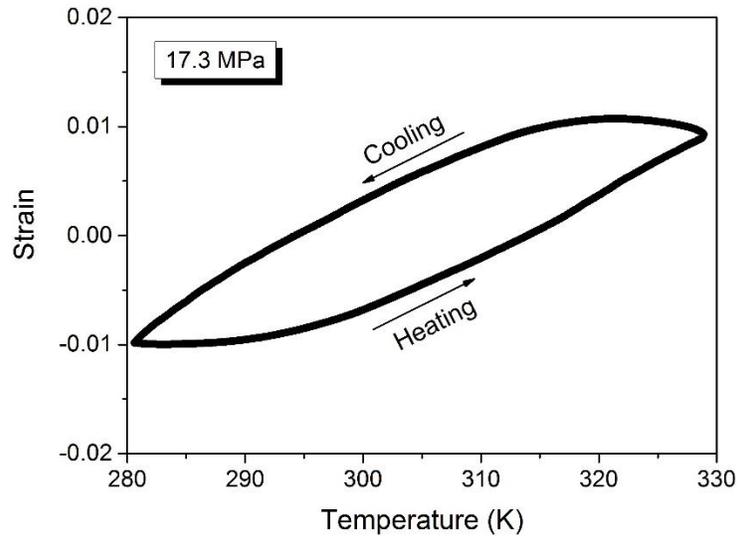

**Figure A2.** Strain *vs.* temperature curves obtained during cooling and heating processes, at 17.3 MPa, according to Materials and Methods.

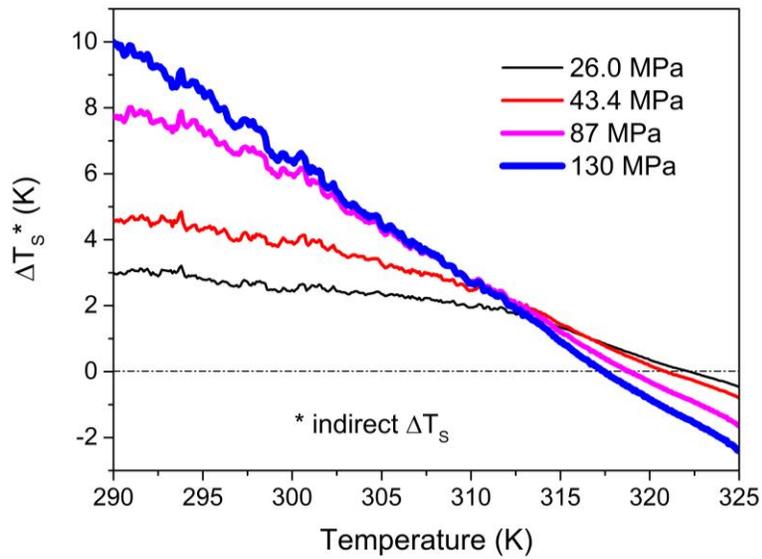

**Figure A3.** Indirect $\Delta T_S$ *vs.* initial temperature, calculated by Eq. 5 using data from Fig. 4a.

**Table A1.** Fitting parameters of $-\Delta T_S$ *vs.* $\sigma_{max}$ curves for PDMS (Fig. 3c), obtained from the power law $|\Delta T_S(T, \sigma_{max})| = a\sigma_{max}^n$.

| Temperature (K) | $a$ (K GPa$^{-n}$) | $n$ |
|---|---|---|
| 243 | 56(4) | 0.88(6) |
| 263 | 62(5) | 0.91(5) |
| 303 | 69(6) | 0.94(6) |

**Table A2:** Fitting parameters of $-\Delta S_T$ *vs.* $\sigma_{max}$ curves for PDMS (Fig. 4b), obtained from the power law $-\Delta S_T(T, \sigma_{max}) = b\sigma_{max}^m$ and from the quadratic function $-\Delta S_T(T, \sigma_{max}) = a_1(T)\sigma_{max} + a_2(T)\sigma_{max}^2$.

| Temperature (K) | $b$ (kJ kg$^{-1}$ K$^{-1}$ GPa$^{-m}$) | $m$ | $a_1$ (kJ kg$^{-1}$ K$^{-1}$ GPa$^{-1}$) | $a_2$ (kJ kg$^{-1}$ K$^{-1}$ GPa$^{-2}$) |
|---|---|---|---|---|
| 290 | 0.25(2) | 0.74(3) | 0.65(1) | -1.9(1) |
| 295 | 0.19(3) | 0.69(7) | 0.622(6) | -2.10(6) |
| 300 | 0.12(3) | 0.6(1) | 0.570(7) | -2.40(8) |
| 305 | 0.05(1) | 0.38(7) | 0.51(3) | -2.6(3) |
| 310 | 0.021(4) | 0.19(7) | 0.40(4) | -2.4(4) |
| 315 | - | - | 0.25(4) | -1.7(4) |
| 320 | - | - | 0.06(2) | -0.8(2) |
| 325 | -0.095(6) | 1.01(3) | -0.091(3) | -0.02(3) |

**Derivation of the expression for ΔS(T,σ) from a modified Landau's theory of elasticity:**

Let us regard the Helmholtz free energy per unit volume as the following series expansion, based on the free energy [44]:

$$F(T, \varepsilon_{ij}) = F_0(T) - B\alpha(T - T_0)\varepsilon_{kk} - \frac{1}{2}B[\beta(T - T_0)^2 - 1]\varepsilon_{kk}^2$$
$$+ G\left(\varepsilon_{ij} - \frac{1}{3}\varepsilon_{kk}\delta_{ij}\right)^2$$

where $F_0$ is the free energy of the unstrained sample; $\alpha$ is the thermal expansion coefficient; $\beta$ accounts for a non-linear thermal deformation of the sample; $B$ and $G$ are the bulk and shear moduli, respectively; $\delta_{ij}$ is the unit tensor; $T_0$ is a reference temperature where the sample experiences no thermal deformation. The expansion above converts the components of a rank-two tensor (the strain tensor $\varepsilon_{ij}$) into a scalar.

It is possible to obtain the entropy through the derivative of the free energy with respect to temperature:

$$S = -\frac{\partial F}{\partial T} = S_0 + B\alpha\varepsilon_{kk} + B\beta(T - T_0)\varepsilon_{kk}^2 \ . \tag{A1}$$

On the other hand, the internal stress is obtained differentiating the free energy with respect to the strain:

$$\sigma_{ij} = \frac{\partial F}{\partial \varepsilon_{ij}} = -B\alpha(T - T_0)\delta_{ij} - B\beta(T - T_0)^2 \varepsilon_{kk}\delta_{ij}$$
$$+ B\varepsilon_{kk}\delta_{ij} + 2G\left(\varepsilon_{ij} - \frac{1}{3}\varepsilon_{kk}\delta_{ij}\right) \tag{A2}$$

Let us now consider the case of confined compression by a uniaxial stress, and let us assume that the stress is applied along the $z$ axis. Therefore, the only non-vanishing component of the strain tensor is $\varepsilon_{zz}$. From Eq. A2, the component $\sigma_{zz}$ is:

$$\sigma_{zz} = -B\alpha(T-T_0) + \left\{B[1-\beta(T-T_0)^2] + \tfrac{4}{3}G\right\}\varepsilon_{zz} \tag{A3}$$

$$\varepsilon_{zz} = \frac{1}{A_1}[\sigma_{zz} + B\alpha(T-T_0)] \tag{A4}$$

$$A_1 \equiv B[1-\beta(T-T_0)^2] + \tfrac{4}{3}G$$

Combining Eqs. A1 and A4, the entropy is given by:

$$S = S_0 + \frac{B^2\alpha^2(T-T_0)}{A_1}\left[1 + \frac{B\beta(T-T_0)^2}{A_1}\right]$$
$$+ \frac{B\alpha}{A_1}\left[1 + \frac{2B\beta(T-T_0)^2}{A_1}\right]\sigma_{zz} + \frac{B\beta(T-T_0)}{A_1^2}\sigma_{zz}^2. \tag{A5}$$

Therefore, the entropy change can be expressed as a sum of powers of the applied compressive stress:

$$\Delta S(T, \Delta\sigma) = a_1(T)\Delta\sigma + a_2(T)(\Delta\sigma)^2,$$

where $\Delta\sigma = \sigma_1 - \sigma_0$. When $\sigma_0 = 0$, we may write:

$$\Delta S(T, \sigma) = a_1(T)\sigma + a_2(T)\sigma^2$$

**Satisfying the isostatic condition**

The uniaxial load exerted by the piston is transferred to the confined elastomeric sample. According to Landau's theory of elasticity [44], for homogeneous deformations, a uniaxial load ($\sigma_{zz}$) on a material will lead to transverse stresses ($\sigma_{xx}$ and $\sigma_{yy}$), as follows:

$$\sigma_{xx} = \sigma_{yy} = \left(\frac{3B - 2G}{3B + 4G}\right)\sigma_{zz}$$

where $B$ is the bulk modulus and $G$ is the shear modulus. The isostatic condition is given by: $\sigma_{xx} = \sigma_{yy} = \sigma_{zz}$. For PDMS, $B \approx 3.4\times10^9$ Pa, $G \approx 6.8\times10^5$ Pa, according to Johnston and co-authors [45], and then $\sigma_{xx} = \sigma_{yy} = 0.9996\sigma_{zz}$. Therefore, the applied tension is isostatic and we can name the effect as barocaloric.

## References

[1]  J.A. Gough, A Description of a property of Caoutchouc or Indian Rubber, Mem. Lit. Phyiosophical Soc. Manchester. 1 (1805) 288–295. http://www.biodiversitylibrary.org/bibliography/49075.

[2]  J.P. Joule, On some thermodynamic properties of solids, Phil. Trans. R. Soc. Lond. 149 (1859) 91–131.

[3]  W. Thomson, II. On the thermoelastic, thermomagnetic, and pyroelectric properties of matter, Philos. Mag. Ser. 5. 5 (1878) 4–27. doi:10.1080/14786447808639378.

[4]  V.K. Pecharsky, K.A. Gschneidner, Jr., Giant magnetocaloric effect in $Gd_5(Si_2Ge_2)$, Phys. Rev. Lett. 78 (1997) 4494–4497. doi:10.1103/PhysRevLett.78.4494.

[5]  A.S. Mischenko, K. Zhang, J.F. Scott, R.W. Whatmore, N.D. Mathur, Giant Electrocaloric Effect in Thin-Film $PbZr_{0.95}0Ti_{0.05}O_3$, Science. 311 (2006) 1270–1271. doi:10.1126/science.1123811.

[6]  S.L. Dart, R.L. Anthony, E. Guth, Rise of temperature on fast stretching of synthetic and natural rubbers, Ind. Eng. Chem. 34 (1942) 1340–1342. doi:10.5254/1.3546779.

[7]  S.L. Dart, E. Guth, Rise of temperature on fast stretching of butyl rubber, Rubber Chem. Technol. 18 (1945) 803–816. doi:10.5254/1.3546779.

[8]  E.L. Rodriguez, F.E. Filisko, Thermoelastic temperature changes in poly(methyl methacrylate) at high hydrostatic pressure: Experimental, J. Appl. Phys. 53 (1982) 6536. doi:10.1063/1.330081.

[9]  E. Bonnot, R. Romero, L. Mañosa, E. Vives, A. Planes, Elastocaloric effect associated with the martensitic transition in shape-memory alloys, Phys. Rev. Lett. 100 (2008) 125901. doi:10.1103/PhysRevLett.100.125901.

[10] L. Mañosa, D. González-alonso, A. Planes, E. Bonnot, M. Barrio, J. Tamarit, et al., Giant

solid-state barocaloric effect in the Ni-Mn-In magnetic shape-memory alloy, Nat. Mater. 9 (2010) 478–481. doi:10.1038/nmat2731.

[11] R. Millán-Solsona, E. Stern-Taulats, E. Vives, A. Planes, J. Sharma, A.K. Nayak, et al., Large entropy change associated with the elastocaloric effect in polycrystalline Ni-Mn-Sb-Co magnetic shape memory alloys, Appl. Phys. Lett. 105 (2014) 241901. doi:10.1063/1.4904419.

[12] E. Stern-Taulats, A. Planes, P. Lloveras, M. Barrio, J.L. Tamarit, S. Pramanick, et al., Barocaloric and magnetocaloric effects in $Fe_{49}Rh_{51}$, Phys. Rev. B - Condens. Matter Mater. Phys. 89 (2014) 214105. doi:10.1103/PhysRevB.89.214105.

[13] Y. Liu, J. Wei, P.E. Janolin, I.C. Infante, X. Lou, B. Dkhil, Giant room-temperature barocaloric effect and pressure-mediated electrocaloric effect in BaTiO3 single crystal, Appl. Phys. Lett. 104 (2014) 162904. doi:10.1063/1.4873162.

[14] P. Lloveras, E. Stern-Taulats, M. Barrio, J.-L. Tamarit, S. Crossley, W. Li, et al., Giant barocaloric effects at low pressure in ferrielectric ammonium sulphate, Nat. Commun. 6 (2015) 8801. doi:10.1038/ncomms9801.

[15] J.M. Bermúdez-García, M. Sánchez-Andújar, M.A. Señarís-Rodríguez, A New Playground for Organic-Inorganic Hybrids: Barocaloric Materials for Pressure-Induced Solid-State Cooling, J. Phys. Chem. Lett. 8 (2017) 4419–4423. doi:10.1021/acs.jpclett.7b01845.

[16] J. Tušek, K. Engelbrecht, D. Eriksen, S. Dall'Olio, J. Tušek, N. Pryds, A regenerative elastocaloric heat pump, Nat. Energy. 1 (2016) 16134. doi:10.1038/nenergy.2016.134.

[17] D. Guyomar, Y. Li, G. Sebald, P. Cottinet, B. Ducharne, J. Capsal, Elastocaloric modeling of natural rubber, Appl. Therm. Eng. 57 (2013) 33–38. doi:10.1016/j.applthermaleng.2013.03.032.

[18] Z. Xie, G. Sebald, D. Guyomar, Z. Xie, G. Sebald, D. Guyomar, Elastocaloric effect dependence on pre-elongation in natural rubber Elastocaloric, Appl. Phys. Lett. 107 (2015) 81905. doi:10.1063/1.4929395.

[19] T. Matsuo, N. Azuma, Y. Toriyama, T. Yoshioka, Mechanocaloric properties of poly ( dimethylsiloxane ) and ethylene – propylene rubbers, J. Therm. Anal. Calorim. 123 (2016)

1817–1824. doi:10.1007/s10973-015-4675-0.

[20]  S. Patel, A. Chauhan, R. Vaish, P. Thomas, Elastocaloric and barocaloric effects in polyvinylidene di-fluoride-based polymers, Appl. Phys. Lett. 108 (2016) 72903. doi:10.1063/1.4942000.

[21]  Z. Xie, G. Sebald, D. Guyomar, Comparison of direct and indirect measurement of the elastocaloric effect in natural rubber, Appl. Phys. Lett. 108 (2016) 41901. doi:10.1063/1.4940378.

[22]  Y. Yoshida, K. Yuse, D. Guyomar, J. Capsal, G. Sebald, Y. Yoshida, et al., Elastocaloric effect in poly (vinylidene fluoride-trifluoroethylene- chlorotrifluoroethylene) terpolymer, Appl. Phys. Lett. 108 (2016) 242904. doi:10.1063/1.4953770.

[23]  E.O. Usuda, N.M. Bom, A.M.G. Carvalho, Large barocaloric effects at low pressures in natural rubber, Eur. Polym. J. 92 (2017) 287–293. https://arxiv.org/abs/1701.05237.

[24]  G. Sebald, Z. Xie, D. Guyomar, Fatigue effect of elastocaloric properties in natural rubber, Philos. Trans. R. Soc. London A Math. Phys. Eng. Sci. 374 (2016) 439–450. doi:10.1098/rsta.2015.0302.

[25]  J.E. Mark, Polymer Data Handbook, Oxford University Press, New York, EUA, 1999. doi:10.1021/ja907879q.

[26]  L.R.G. Treloar, The physics of rubber elasticity, Oxford University Press, London, 1975. doi:10.1016/0022-3697(59)90114-3.

[27]  M. Rezakazemi, A. Vatani, T. Mohammadi, Synergistic interaction between POSS and fumed silica on the properties of corsslinked PDMS nanocomposite membranes, RSC Adv. 5 (2015) 82460–82470.

[28]  F.L. Pissetti, P.L. De Araújo, F.A.B. Silvaa, G.Y. Poirier, Synthesis of poly(dimethylsiloxane) networks functionalized with imidazole or benzimidazole for copper(II) removal from water, J. Braz. Chem. Soc. 26 (2015) 266–272. doi:10.5935/0103-5053.20140264.

[29]  N.M. Bom, E.O. Usuda, G.M. Guimarães, A.A. Coelho, A.M.G. Carvalho, Note: Experimental setup for measuring the barocaloric effect in polymers: Application to

natural rubber, Rev. Sci. Instrum. 88 (2017) 46103. http://arxiv.org/abs/1612.08638.

[30] K.A. Gschneidner, V.K. Pecharsky, Magnetocaloric Materials, Annu. Rev. Mater. Sci. 30 (2000) 387–429.

[31] X. Moya, S. Kar-Narayan, N.D. Mathur, Caloric materials near ferroic phase transitions, Nat. Mater. 13 (2014) 439–450. doi:10.1038/nmat3951.

[32] B. Lu, J. Liu, Mechanocaloric materials for solid-state cooling, Sci. Bull. 60 (2015) 1638. doi:10.1007/s11434-015-0898-5.

[33] L. Manosa, A. Planes, Mechanocaloric effects in Shape Memory Alloys, Philos. Trans. R. Soc. London A Math. Phys. Eng. Sci. 374 (2016) 20150310. doi:10.1098/rsta.2015.0310.

[34] M.M. Vopson, Theory of giant-caloric effects in multiferroic materials, J. Phys. D. Appl. Phys. 46 (2013) 345304. doi:10.1088/0022-3727/46/34/345304.

[35] Y. Liu, I.C. Infante, X. Lou, L. Bellaiche, J.F. Scott, B. Dkhil, Giant room-temperature elastocaloric effect in ferroelectric ultrathin films, Adv. Mater. 26 (2014) 6132–6137. doi:10.1002/adma.201401935.

[36] A.M.G. Carvalho, A.A. Coelho, S. Gama, P.J. von Ranke, C.S. Alves, Isothermal variation of the entropy (delta S) for the compound $Gd_5Ge_4$ under hydrostatic pressure, J. Appl. Phys. 104 (2008) 63915. doi:10.1063/1.2980040.

[37] *A. M. Tishin and Y. I. Spichkin, "The Magnetocaloric Effect and its Applications", 1st edition (Institute of Physics, Bristol and Philadelphia, 2003).*

[38] B. Wang, S. Krause, Properties of Dimethylsiloxane Microphases in Phase-Separated Dimethylsiloxane Block Copolymers, Macromolecules. 20 (1987) 2201–2208. doi:10.1021/ma00175a026.

[39] J.C. Mitchell, D.J. Meier, Rapid Stress-Induced Crystallization in Natural Rubber, J. Polym. Sci. Part A-2 Polym. Phys. 6 (1968) 1689–1703.

[40] S. Yuce, M. Barrio, B. Emre, E. Stern-Taulats, A. Planes, J.L. Tamarit, et al., Barocaloric effect in the magnetocaloric prototype $Gd_5Si_2Ge_2$, Appl. Phys. Lett. 101 (2012) 71906. doi:10.1063/1.4745920.

[41] D. Matsunami, A. Fujita, K. Takenaka, M. Kano, Giant barocaloric effect enhanced by the

frustration of the antiferromagnetic phase in Mn$_3$GaN, Nat. Mater. 14 (2015) 73–78. doi:10.1038/NMAT4117.

[42] E. Stern-Taulats, A. Gràcia-Condal, A. Planes, P. Lloveras, M. Barrio, J.L. Tamarit, et al., Reversible adiabatic temperature changes at the magnetocaloric and barocaloric effects in Fe$_{49}$Rh$_{51}$, Appl. Phys. Lett. 107 (2015) 152409. doi:10.1063/1.4933409.

[43] J.M. Bermúdez-García, M. Sánchez-Andújar, S. Castro-García, J. López-Beceiro, R. Artiaga, M.A. Señarís-Rodríguez, Giant barocaloric effect in the ferroic organic-inorganic hybrid [TPrA][Mn(dca)$_3$] perovskite under easily accessible pressures, Nat. Commun. 8 (2017) 15715. doi:10.1038/ncomms15715.

[44] L.D. Landau, E.M. Lifshitz, Theory of Elasticity: Vol. 7 of Course of Theoretical Physics, Pergamon Press, Oxford, 1970. doi:10.1063/1.3057037.

[45] I.D. Johnston, D.K. McCluskey, C.K.L. Tan, M.C. Tracey, Mechanical characterization of bulk Sylgard 184 for microfluidics and microengineering, J. Micromechanics Microengineering. 24 (2014) 35017. doi:10.1088/0960-1317/24/3/035017.